\documentclass[twocolumn]{aa}
\usepackage{graphicx}
\usepackage{txfonts}
\usepackage{natbib}
\bibpunct{(}{)}{;}{a}{}{,}

\begin{document}

\title{ The HARPS search for southern extra-solar planets\thanks{Based
    on observations made with the HARPS instrument on the ESO 3.6 m
    telescope at La Silla Observatory under the GTO programme ID
    072.C-0488.}  } 
\subtitle{V. A 14 Earth-masses planet orbiting HD\,4308}

\author{
S.~Udry\inst{1}
\and M.~Mayor\inst{1}
\and W.~Benz\inst{2}
\and J.-L.~Bertaux\inst{3}
\and F.~Bouchy\inst{4}
\and C.~Lovis\inst{1}
\and C.~Mordasini\inst{2}
\and F.~Pepe\inst{1}
\and D.~Queloz\inst{1}
\and J.-P.~Sivan\inst{4}
}

\offprints{S. Udry}

\institute{
Observatoire de Gen\`eve, 51 ch. des Maillettes, 1290 Sauverny, 
Switzerland\\
\email{stephane.udry@obs.unige.ch}
\and
Physikalisches Institut Universit\"at Bern, Sidlerstrasse 5, 3012
Bern, Switzerland
\and
Service d'A\'eronomie du CNRS, BP 3, 91371 Verri\`eres-le-Buisson,
France 
\and
Laboratoire d'Astrophysique de Marseille, Traverse du Siphon, 13013
Marseille, France
}

\date{Received: 22.08.2005 ; accepted: 28.09.2005}

\abstract{We present here the discovery and characterisation of a very
  light planet around HD\,4308.  The planet orbits its star in
  15.56\,days.  The circular radial-velocity variation presents a tiny
  semi-amplitude of 4.1 ms$^{-1}$ that corresponds to a planetary
  minimum mass $m_2\sin{i}$\,=\,14.1\,M$_{\oplus}$ (Earth masses).
  The planet was unveiled by high-precision radial-velocity
  measurements obtained with the HARPS spectrograph on the ESO 3.6-m
  telescope. The radial-velocity residuals around the Keplerian
  solution are 1.3\,ms$^{-1}$, demonstrating the very high quality of
  the HARPS measurements.  Activity and bisector indicators exclude
  any significant perturbations of stellar intrinsic origin, which
  supports the planetary interpretation.  Contrary to most planet-host
  stars, HD\,4308 has a marked sub-solar metallicity
  ([Fe/H]\,=\,$-$0.31), raising the possibility that very light planet
  occurrence might show a different coupling with the parent star's
  metallicity than do giant gaseous extra-solar planets. Together with
  Neptune-mass planets close to their parent stars, the new planet
  occupies a position in the mass-separation parameter space that is
  constraining for planet-formation and evolution theories.  The
  question of whether they can be considered as residuals of
  evaporated gaseous giant planets, ice giants, or super-earth planets
  is discussed in the context of the latest core-accretion models.

 }

\maketitle

\section{Introduction}

After a decade of enthusiastic discoveries in the field of extra-solar
gaseous giant planets, mainly coming from large high-precision
radial-velocity surveys of solar-type stars, the {\sl quest for other
  worlds} has now passed a new barrier with the recent detections of
several planets in the Neptune-mass regime
\citep{McArthur-2004,Santos-2004:a,Butler-2004,Vogt-2005,Rivera-2005,Bonfils-2005}.
They are supposedly mainly composed of icy/rocky material, being
formed without or having lost the extended gaseous atmosphere expected
to grow during the planet migration towards the centre of the system.

This new step forward has been made possible primarily thanks to the
development of a new generation of instruments capable of
radial-velocity measurements of unprecedented quality. The ``fer de
lance'' among them is undoubtedly the ESO high-resolution HARPS
fiber-fed echelle spectrograph especially designed for planet-search
programmes and asteroseismology. HARPS has already proved to be the
most precise spectro-velocimeter to date, as it reaches an
instrumental radial-velocity accuracy at the level of 1\,ms$^{-1}$
over months/years \citep{Mayor-2003:a,Lovis-2005}. The precision
achieved is even better on a short-term basis \citep{Bouchy-2005}.
Another fundamental change that has allowed this progress in planet
detection towards the very low masses is the application of a careful
observing strategy to reduce as much as possible the perturbing effect
of stellar oscillations hiding the tiny radial-velocity signal induced
on solar-type stars by Neptune-mass planets. This is discussed further
in detail in Sect.\,\ref{SectMes}.

\begin{figure}
\centering
\includegraphics[width=\hsize]{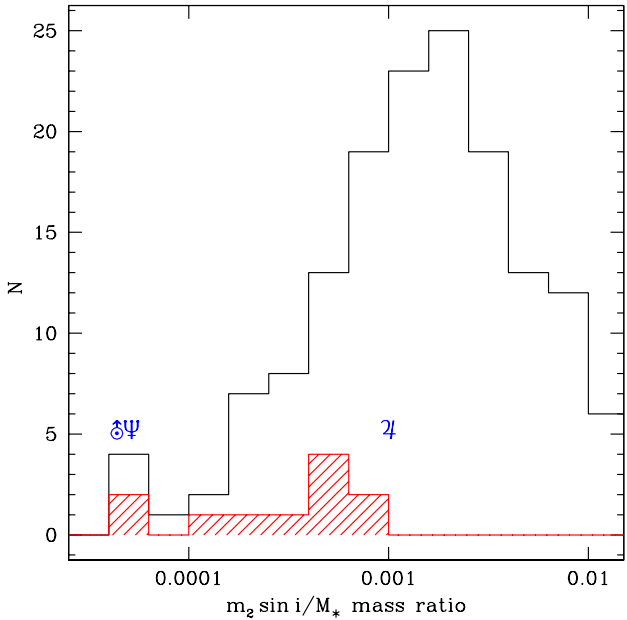}
\caption{Distribution of planet to primary-star mass ratios for the 
  known exoplanets and low-mass brown dwarfs. The HARPS detections are
  represented by the hatched histogram. Planet symbols indicate the
  positions of Jupiter-, Neptune-, and Uranus-mass objects.}
\label{Figfm}
\end{figure}

In about 1.5 years, the GTO HARPS planet-search programme has
discovered (or co-discovered) 11 planets, amongst which there are: 5
hot Jupiters \citep{Pepe-2004,Moutou-2005,LoCurto-2005}; 3 planets
with masses in the sub-Saturn mass regime and orbiting at moderate
distances from their parent stars \citep{Lovis-2005}; and 3
Neptune-mass planets, namely a 14.4\,M$_{\oplus}$ planet around
$\mu$\,Ara \citep{Santos-2004:a}, a 16.6\,M$_{\oplus}$ planet around
the M3 Gl\,581 \citep{Bonfils-2005}, and the planet described in this
paper.  As illustrated in Fig\,\ref{Figfm}, these planets lie mainly
in the low-mass tail of the planetary-mass distribution. This
distribution, which was known to be biased towards low masses, can now
be explored in greater detail.  The characterisation of the low-mass
objects will strongly constrain planet-formation and evolution
scenarios, as planets with masses between 10 and 100\,M$_{\oplus}$ are
not expected in large numbers according to some current formation
models \citep[see e.g. ][]{Ida-2004:a}.

The discovery of the extremely low-mass planets represents a new
benchmark for planet surveys and provides information on the low end
of the planetary-mass distribution.  In this paper, we present the
discovery of a new very low-mass companion to the star HD\,4308.
Interesting characteristics of this planet include its Uranus-like
mass and the low metallicity of its parent star.  The paper is
structured as follows.  Section\,2 briefly describes the physical
properties of the parent star. The radial-velocity measurements,
observation strategy, and orbital solution are presented in Sects.\,3
and 4. Finally, we discuss the characteristics of the new planet in
Sect.\,5 in the context of the latest core-accretion planet-formation
models, whereas the last section is devoted to some concluding
remarks.

\section{Stellar characteristics of HD\,4308}

The basic photometric (G5V, $V$\,=\,6.54, $B$\,$-$\,$V$\,=\,0.65) and
astrometric ($\pi$\,=\,45.76\,mas) properties of HD\,4308 were taken
from the Hipparcos catalogue \citep{Esa-97}. They are recalled in
Table\,\ref{TableStar}, together with inferred quantities like the
absolute magnitude ($M_V$\,=\,4.85) and the stellar physical
characteristics derived from the HARPS spectra following the method
described in \citet{Santos-2001:a,Santos-2004:b,Santos-2005}.  A
standard local thermodynamical equilibrium (LTE) analysis was applied
to a {\sl resulting} spectrum built as the sum of the 205 individual
spectra gathered for the star (see next section).  On this resulting
spectrum, we measured an equivalent S/N of $\sim$\,1100 in the Li
wavelength region (6700\,$\AA$). This study thus provides very precise
values for the effective temperature ($T_{\rm
eff}$\,=\,5686\,$\pm$\,13), metallicity
([Fe/H]\,=\,$-0.31$\,$\pm$\,0.01), and surface gravity
($\log{g}$\,=\,4.49\,$\pm$\,0.07) of the star. The quoted
uncertainties do not include systematic errors, such as the use of
different temperature scales.  However, these systematic errors should
be small, in particular the ones concerning stellar metallicity
\citep[see discussion in][ 2005]{Santos-2004:b}.

Very interestingly, HD\,4308 has a sub-solar metallicity contrary to
most planet-host stars.  The high values of the spatial velocities of
the star ($U$\,=\,$-52$\,kms$^{-1}$; $V$\,=\,$-110$\,kms$^{-1}$;
$W$\,=\,$-29$\,kms$^{-1}$) further points towards the star probably
belonging to the thick disk.

\begin{figure}
\centering
\includegraphics[width=0.7\hsize,angle=270.]{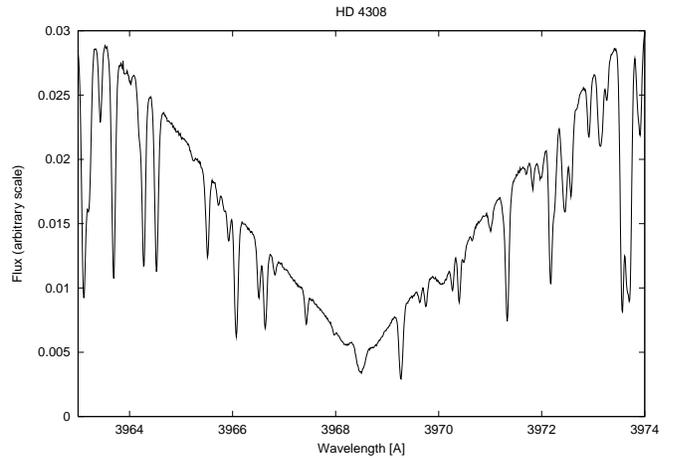}
\caption{\ion{Ca}{ii}\,H ($\lambda$ = 3968.47 \AA) absorption line
  region of the summed HARPS spectra for HD\,4308. Re-emission in the
  core of the line is absent, showing the low chromospheric activity
  of the star.}
\label{FigCaII}
\end{figure}

From the colour index, the derived effective temperature, and the
corresponding bolometric correction, we estimated the star luminosity
as 0.99\,L$_\odot$. Except for its metallicity, the star is thus very
similar to the Sun. Because of the low metallicity, however, we
interpolated a sub-solar mass ($M_\star$\,=\,0.83\,M$_{\odot}$) in the
grid of Geneva stellar evolutionary models with appropriate metal
abundance \citep{Schaerer-93:a}. The age derived from the model also
points towards an old star (age\,$>$\,10\,Gyr).  The projected
rotational velocity $v\sin{i}$\,=\,1.2\,kms$^{-1}$ was estimated using
the calibration of the CORALIE cross-correlation function given in
\citet{Santos-2002}.  Table \ref{TableStar} gathers those values as
well.

\begin{figure}
\centering
\includegraphics[width=\hsize]{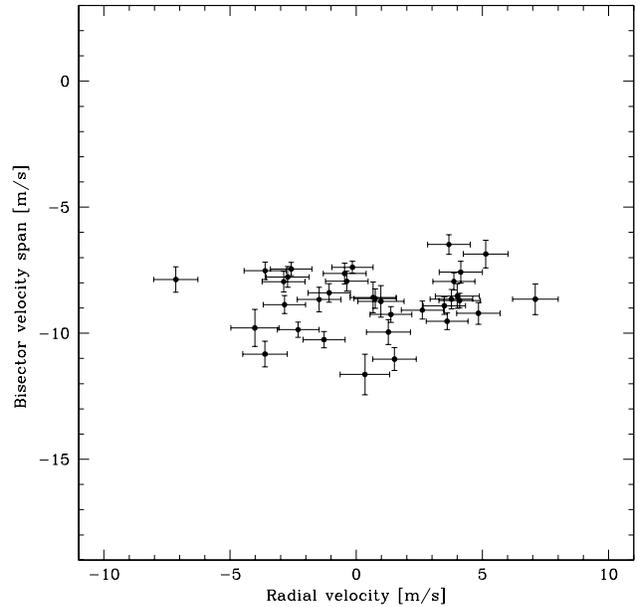}
\caption{Bisector velocity span as a function of the measured radial 
  velocity. No correlation is found supporting a non-intrinsic origin
  for the radial-velocity variation.}
\label{FigBis}
\end{figure}

We also computed the $\log R'_{\mathrm{HK}}$ activity indicator from
the spectra, measuring the re-emission flux in the \ion{Ca}{ii}\,H and
K lines corrected for the photospheric flux contribution.  This index
represents a useful tool for estimating the stellar radial-velocity
jitter expected for the star due to rotational modulation of star
spots or other active regions on the stellar surface
\citep{Saar-97,Santos-2000:b}.  From this indicator we derived a
stellar rotation period $P_{rot}$\,=\,24\,days \cite[following
][]{Noyes-84}, as well as an activity-based age estimate of
$\sim$\,4.3\,Gyr \citep[from the calibration by
][]{Donahue-93,Henry-96}\footnote{Although the relation has been shown
  not to be reliable for ages greater than $\sim$\,2\,Gyr
  \citep{Pace-2004}, it is still helpful to distinguish between
  active, young stars and chromospherically quiet, old stars.}.
HD\,4308 was found to be non-active as shown by the low value of $\log
R'_{\mathrm{HK}}$\,=\,$-4.93$\footnote{A value $\log
  R'_{\mathrm{HK}}$\,=\,$-5.05$ is given in \citet{Henry-96} leading
  to an age estimate of 6.3\,Gyr.} and the absence of re-emission
observed in the core of the CaII\,H line displayed in
Fig.\,\ref{FigCaII}.  Together with the measured small $v \sin{i}$,
these features indicate very low activity-induced radial-velocity
jitter for this star. Moreover, analysis of the bisector shape of the
cross-correlation function \citep[see details on the method in
][]{Queloz-2001:a} shows no variations in the CCF profile down to the
photon noise level (Fig.\,\ref{FigBis}), giving strong support to a
planetary origin of the radial-velocity variations.

\begin{table}
\caption{Observed and inferred parameters of
  HD\,4308. Photometric and astrometric parameters were taken from the
  Hipparcos catalogue \citep{Esa-97}. The other stellar physical
  quantities were obtained from a high-resolution ETL spectral
  analysis or were interpolated in the grid of Geneva evolution models
  \citep{Schaerer-93:a}.}
\label{TableStar}
\centering
\begin{tabular}{l l c c}
\hline\hline
\multicolumn{2}{l}{\bf Parameter} &\hspace*{2mm} & \bf HD\,4308 \\
\hline
Sp & & & G5V  \\
$V$ & [mag] & & 6.54 \\
$B-V$ & [mag] & & 0.65  \\
$\pi$ & [mas] & & 45.76 \\
$M_V$ & [mag] & & 4.85 \\
$T_{\mathrm{eff}}$ & [K] & & 5685 $\pm$ 13 \\
log $g$ & [cgs] & & 4.49 $\pm$ 0.07 \\
$\mathrm{[Fe/H]}$ & [dex] & & $-0.31$ $\pm$ 0.01 \\
$L$ & [$L_{\odot}$] & & 0.99  \\
$M_*$ & [$M_{\odot}$] & & 0.83 $\pm$ 0.01 \\
$v\sin{i}$ & [km s$^{-1}$] & & 1.2 \\
$\log R'_{\mathrm{HK}}$ & & & $-4.93$ \\      
$P_{\mathrm{rot}}$($\log R'_{\mathrm{HK}}$) & [days] & & 24 \\
age(model/$\log R'_{\mathrm{HK}}$) & [Gyr] & & $>$\,10\,/\,4.3 \\
\hline
\end{tabular}
\end{table}

\section{Very high precision radial-velocity measurements and 
  observing strategy}
\label{SectMes}

HD\,4308 is part of the HARPS ``high-precision'' GTO subsample that
aims at detecting very low-mass extra-solar planets by pushing the
radial velocity measurement accuracy below the 1\,ms$^{-1}$ mark.
However, asteroseismology observations carried out by HARPS have made
clear that the precision the instrument is capable of achieving is no
longer set by instrumental characteristics but rather by the stars
themselves \citep{Mayor-2003:a,Bouchy-2005}. Indeed, stellar p-mode
oscillations on short time-scales and stellar jitter on longer
time-scales can and do induce significant radial-velocity changes at
the level of accuracy of HARPS measurements.  For instance, even a
very ``quiet'' G or K dwarf shows oscillation modes of several tens of
cms$^{-1}$ each, which might add up to radial-velocity amplitudes as
large as several ms$^{-1}$ \citep{Bouchy-2005}. As a consequence, any
exposure with a shorter integration time than the oscillation period
of the star, or even than mode-interference variation time-scales,
might fall arbitrarily on a peak or on a valley of these mode
interferences and thus introduce additional radial-velocity
``noise''. This phenomenon could, therefore, seriously compromise the
ability to detect very low-mass planets around solar-type stars by
means of the radial-velocity technique.

To minimize these effects as much as possible, the stars of the HARPS
GTO ``high-precision'' sample have been selected from the CORALIE
planet-search sample \citep{Udry-2000:a} as slowly rotating,
non-evolved, and low-activity stars with no obvious radial-velocity
variations at the CORALIE precision level (typically 10\,ms$^{-1}$ for
HD\,4308). Moreover, in order to average out stellar oscillations, the
observations are designed to last at least 15 minutes on the target,
splitting them in several exposures, when required, to avoid CCD
saturation.  The radial velocities of individual exposures are
obtained with the standard HARPS reduction pipeline, based on the
cross-correlation with an appropriate stellar template, the precise
nightly wavelength calibration with ThAr spectra and the tracking of
instrumental drifts with the simultaneous ThAr technique
\citep{Baranne-96}. The final radial velocity is then obtained by
averaging the multiple consecutive exposures.

\begin{figure}
\centering
\includegraphics[width=\hsize]{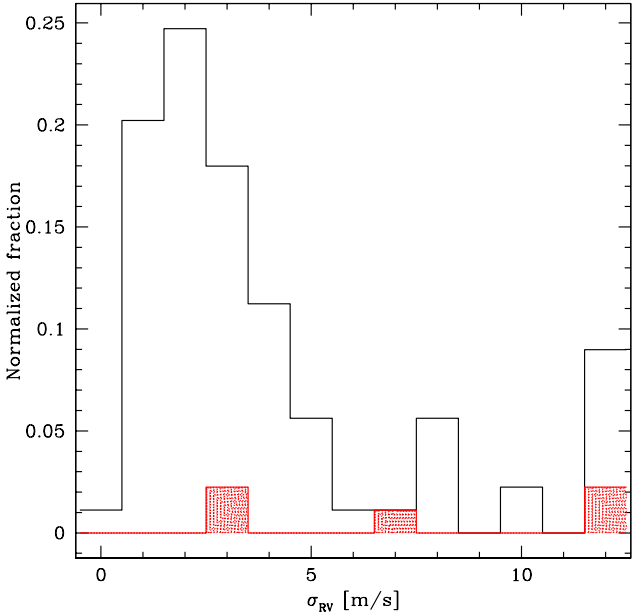}
\caption{Histogram of the observed radial-velocity dispersion ($\sigma_{RV}$) 
  of the stars in the HARPS ``high-precision'' sub-programme.  It has
  been produced using the data obtained with the new observing
  strategy and includes only stars with more than 3 measurements.  The
  position of the planets detected with HARPS is indicated by the
  hatched area.  }
\label{FigSigvr}
\end{figure}

This strategy is now applied to all stars in the ``high-precision''
programme. The results are best summarised by the histogram of the
radial-velocity dispersion of the HARPS measurements for this
programme (Fig.\,\ref{FigSigvr}). The distribution mode is just below
2\,ms$^{-1}$, and the peak decreases rapidly towards higher values.
More than 80\,\% of the stars show smaller dispersion than
5\,ms$^{-1}$, and more than 35\,\% have dispersions below
2\,ms$^{-1}$. It must be noted that the computed dispersion includes
photon-noise error, wavelength-calibration error, stellar oscillations
and jitter, and, in particular, it is ``polluted'' by known extrasolar
planets (hatched part in Fig.\,\ref{FigSigvr}) and still undetected
planetary companions.

\begin{table}[t!]
\caption{Multiple-exposure averaged radial-velocities and
  uncertainties (including calibration errors) for HD\,4308. 
All data are relative to the solar system barycentre.}
\label{TableRV}
\centering
\begin{tabular}{c c c}
\hline\hline
\bf JD-2400000 & \bf RV & \bf Uncertainty \\
\bf [days] & \bf [km\,s$^{-1}$] & \bf [km\,s$^{-1}$] \\
\hline
52899.77052  & 95.24540  & 0.00136 \\
52939.67749  & 95.25140  & 0.00144 \\
52946.72659  & 95.24100  & 0.00157 \\
53266.77155  & 95.24955  & 0.00102 \\
53268.72143  & 95.24940  & 0.00100 \\
53273.74140  & 95.24154  & 0.00100 \\
53274.74846  & 95.24200  & 0.00095 \\
53287.69258  & 95.24542  & 0.00101 \\
53288.69001  & 95.24426  & 0.00088 \\
53289.70387  & 95.24530  & 0.00089 \\
53291.68596  & 95.23983  & 0.00109 \\
53292.67405  & 95.24129  & 0.00092 \\
53294.64135  & 95.24244  & 0.00115 \\
53295.76250  & 95.24850  & 0.00129 \\
53298.72754  & 95.24822  & 0.00090 \\
53308.68463  & 95.24248  & 0.00088 \\
53309.54646  & 95.24317  & 0.00095 \\
53310.65747  & 95.24471  & 0.00088 \\
53312.54143  & 95.24896  & 0.00094 \\
53314.70856  & 95.24910  & 0.00094 \\
53315.62201  & 95.24953  & 0.00094 \\
53321.64168  & 95.23840  & 0.00116 \\
53366.54310  & 95.24383  & 0.00087 \\
53367.58691  & 95.24281  & 0.00088 \\
53371.56598  & 95.24321  & 0.00088 \\
53372.55973  & 95.24643  & 0.00088 \\
53373.55895  & 95.24720  & 0.00094 \\
53374.55931  & 95.24707  & 0.00090 \\
53376.54410  & 95.24964  & 0.00088 \\
53377.53686  & 95.24710  & 0.00089 \\
53401.53108  & 95.24470  & 0.00126 \\
53403.54910  & 95.24422  & 0.00100 \\
53404.52408  & 95.24675  & 0.00099 \\
53406.52597  & 95.24824  & 0.00100 \\
53408.52500  & 95.25076  & 0.00105 \\
53409.52594  & 95.24960  & 0.00101 \\
53410.52003  & 95.24961  & 0.00100 \\
53412.50936  & 95.24525  & 0.00101 \\
53573.91705  & 95.24330  & 0.00112 \\
53576.87555  & 95.24970  & 0.00111 \\
53579.83685  & 95.24990  & 0.00115 \\
\hline
\end{tabular}
\end{table}

\section{Orbital Keplerian solution for HD\,4308\,b}

A set of 205 individual high S/N spectra covering 680~days were
gathered for HD\,4308, as described in Sect\,\ref{SectMes}.
They finally corresponded to 41 radial-velocity measurements after the
multiple-exposure averages.  Typical exposure times of 5 minutes
yielded S/N ratios of $\sim$\,150 per pixel at 550 nm for individual
spectra, corresponding to individual photon-noise errors of
$\sim$\,0.5\,ms$^{-1}$. After combining the spectra, we quadratically
added the night wavelength calibration error\footnote{Presently on the
  order of 80\,cms$^{-1}$, this error represents the main limitation
  on the measurement precision, but it will improve soon thanks to an
  ongoing redefinition of the thorium lines used for the wavelength
  calibration. See also \citet{Lovis-2005} for further details on the
  different error sources of the radial-velocity estimate}. The list
of the averaged radial-velocity measurements obtained for HD\,4308 is
given in Table\,\ref{TableRV}.

\begin{figure}[t]
\centering
\includegraphics[width=\hsize]{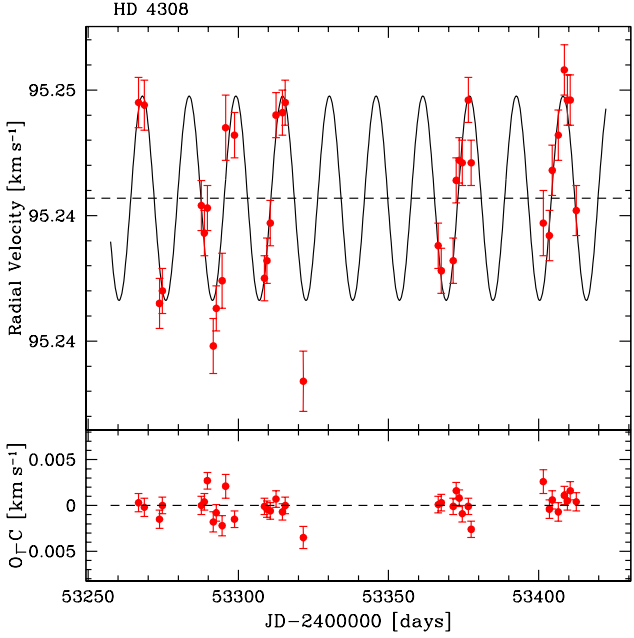}
\caption{Intermediate season of HARPS radial velocities for HD\,4308. 
  The best Keplerian fit to the data (solid curve) gives a minimum
  mass of 14.1 $M_{\oplus}$ and an orbital period of 15.56 days
  for the planet.}
\label{Figdvft}
\end{figure}

\begin{figure}[t]
\centering
\includegraphics[width=\hsize]{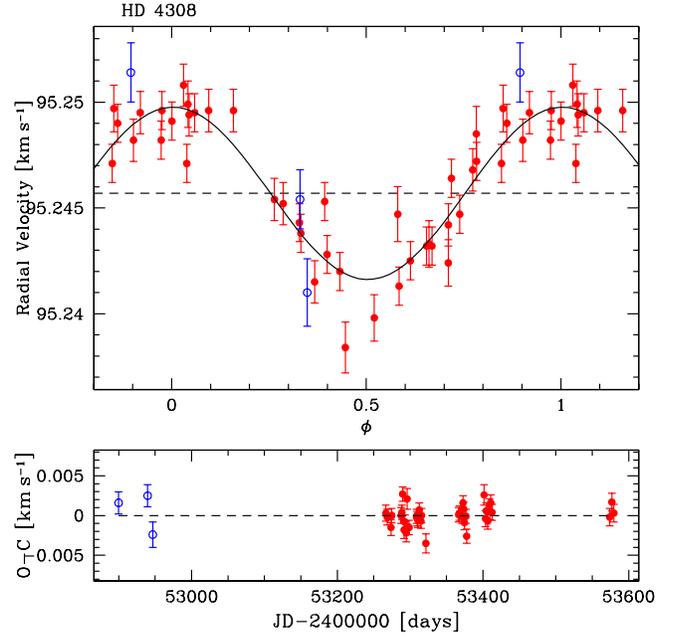}
\caption{Phase-folded radial velocities of HD\,4308 superimposed
  on the best Keplerian orbital solution (solid curve).  The 3 older
  points obtained before introducting the new observing
  strategy (see text) are indicated by open symbols.}
\label{Figphas}
\end{figure}

The intermediate-season velocities are displayed in
Fig.\,\ref{Figdvft} with the best Keplerian solution fitted to the
data. The derived period is 15.56~days, and the radial-velocity
semi-amplitude $K$ is 4.1\,ms$^{-1}$.  Despite the small amplitude of
the radial-velocity variation, we estimated a small false-alarm
probability for the period derived of 10$^{-3}$ from Monte-Carlo
simulations.  The best Keplerian planetary solution is circular.
Considering a primary mass of 0.83\,M$_{\odot}$, these parameters lead
to a minimum mass
$m_2\sin{i}$\,=\,0.0442\,M$_{\mathrm{Jup}}$\,=\,14.1\,M$_{\oplus}$ and
a separation $a$\,=\,0.115\,AU for the planet.  Figure\,\ref{Figphas}
shows the radial velocities folded to the orbital period, including 3
older ``single-shot'' measurements obtained before the change in our
observational strategy. Orbital and inferred parameters for the
non-circular solution are given in Table\,\ref{TabOrb}.

The weighted rms of the residuals around the solution is
1.3\,ms$^{-1}$, which is larger than the internal errors; but, as in
the case of $\mu$\,Ara \citep{Santos-2004:a}, they are most probably
due to residuals of unaveraged stellar oscillation modes
\citep{Bouchy-2005}. No additional trend of the radial velocities is
expected, as CORALIE measurements show that the star is constant at a
10\,ms$^{-1}$ level over close to 5~years.

\begin{table}
\caption{Orbital and physical parameters for HD\,4308\,b}
\label{TabOrb}
\centering
\begin{tabular}{l l l c}
\hline\hline
\multicolumn{2}{l}{\bf Parameter} &\hspace*{2mm} & \bf HD\,4308\,b \\
\hline
$P$ & [days] & & 15.56 $\pm$ 0.02 \\
$T$ & [JD-2400000] & & 53314.70 $\pm$ 2.0 \\
$e$ & & & 0.00 $+$ 0.01  \\
$V$ & [km s$^{-1}$] & & 95.2457 $\pm$ 0.0002  \\
$\omega$ & [deg] & & 359 $\pm$ 47 \\
$K$ & [m s$^{-1}$] & & 4.07 $\pm$ 0.2 \\
$a_1 \sin{i}$ & [10$^{-6}$ AU] & & 5.826  \\
$f(m)$ & [10$^{-13} M_{\odot}$] & & 1.09 \\
$m_2 \sin{i}$ & [$M_{\mathrm{Jup}}$] & & 0.0442 \\
$a$ & [AU] & & 0.115\\
\hline
$N_{\mathrm{meas}}$ & & & 41 \\
{\it Span} & [days] & & 680 \\
$\sigma$ (O-C) & [ms$^{-1}$] & & 1.3 \\
$\chi^2_{\rm red}$ & & & 1.92 \\
\hline
\end{tabular}
\end{table}

\section{Discussion}

\subsection{Constraining formation and evolution scenarios}

Low-mass giant planets, i.e. planets with masses in the range
10-100\,M$_{\oplus}$, are particularly interesting as they provide
potentially strong constraints on current models of giant planet
formation and evolution.  Indeed, and perhaps contrary to intuition,
the formation of these objects within the current theoretical models
appears more difficult than the formation of their more massive
counterparts. For this reason, objects with masses within or at the
edge of this range, like $\mu$\,Ara\,c and HD\,4308\,b, are especially
interesting.

In the direct collapse scenario, planets form through gravitational
collapse of patches of the proto-planetary disk \citep{Boss-2002} on
very short timescales. High-resolution simulations
\citep{Mayer-2002,Mayer-2004} of this process show that planets tend
to form on elliptical orbits with a semi-major axis of several
astronomical units and masses between 1 and 7\,M$_{\rm Jup}$.  In this
scenario, $\mu$\,Ara-type planets would have to result from the
subsequent evolution involving migration and from a very significant
amount of mass loss.

In the framework of the core accretion model \citep[e.g.
][]{Pollack-96}, the final mass of a planet is actually determined by
the amount of gas the core accretes after it has reached a critical
mass, which is about 10-15\,M$_{\oplus}$. In \citet{Ida-2004:a},
this amount is determined by the rate at which gas can be accreted
(essentially the Kelvin-Helmholtz timescale) and by the total amount
of gas available within the planet's gravitational reach. Since for
super-critical cores the Kelvin-Helmoltz timescale is short and the
amount of gas available large (compared to an Earth's mass) even in
low mass disks, planets tend to form either less or significantly more
massive than the critical mass. From a large number of formation model
calculations, \citet{Ida-2004:a} have found that only very few planets
form in the mass range 10-100\,M$_{\oplus}$, a range they actually
call \textit{a planetary desert}.  

In the extended core accretion models by \cite{Alibert-2005}, the
planet should be able to accrete gas over the entire lifetime of the
disk, due to its migration. However, since the latter thins out with
time and the planet eventually opens a gap as it grows more massive,
the gas supply decreases with time. The growth rate of the planet is
actually set by the rate at which the disk can supply the gas rather
than by the rate at which the planet can accrete it. Monte Carlo
simulations are ongoing to check whether these models lead to a
different planetary initial mass function than in \citet{Ida-2004:a}.

At first glance, the relatively numerous small-mass objects discovered
so far seem to pose a problem for current planet formation theories
\citep{Lovis-2005}; however, the situation is actually more complex.
For example, examining a potential metallicity-mass relation for
exoplanets, \citet{Rice-2005} raise the possibility that the planet
desert might be populated by late-forming planets for which the
evolution/migration is stopped at intermediate masses and distances by
the dissolution of the disk.  Also, since all the very low-mass
planets that are currently known are located close to their star, one
cannot exclude the possibility that these objects were initially much
more massive and then have lost a significant amount of their mass
through evaporation during their lifetime \citep[see ][ for a more
detailed discussion]{Baraffe-2004,Baraffe-2005}.

While mass loss from initially more massive objects could possibly
account for the light planets very close to their star, it is not
clear whether $\mu$\,Ara\,c located at a distance of 0.09\,AU could
actually result from the evaporation of a more massive object. The
situation is even more critical for HD\,4308\,b that is located
farther away (0.115\,AU) from its parent star, which is, in addition,
less luminous than $\mu$\,Ara by a factor of $\sim$\,1.8. The effects
could possibly be compensated for, at least partially, by the
estimated very old age of the star. However, as more $\mu$\,Ara-like
objects are being discovered, and if they are all the results of the
evaporation of larger mass planets, the question of the probability of
detecting these systems shortly before complete evaporation will
become a central one.  As already stated by \citet{Baraffe-2005}, the
current evaporation models are still affected by large uncertainties
-- lack of detailed chemistry, non-standard chemical composition in
the envelope, effect or rocky/icy cores, etc. -- that will need to be
clarified in order to solve the question of the possible formation of
$\mu$\,Ara-type planets through evaporation.

Finally, given their close location to their star, the detected small
mass planets are likely to have migrated to their current position
from further out in the nebula.  The chemical composition of these
planets will depend upon the extent of their migration and the thermal
history of the nebula and, hence, on the composition of the
planetesimals along the accretion path of the planet. The situation is
complicated by the fact that the ice-line itself is moving as the
nebula evolves \citep[see e.g. ][]{Sasselov-2000}.  Detailed models of
planetary formation that include these effects have yet to be
developed.

\subsection{Influence of parent star metallicity}

It is well-established that the detected giant planets preferentially
orbit metal-rich stars
\citep{Gonzalez-1998:a,Gonzalez-1998:b,Gonzalez-2001,Santos-2001:a,Santos-2003:a,Santos-2005,Fischer-2005}.
The frequency of planets is even found to be a steeply rising function
of the parent star's metallicity, as soon as the latter is over solar
\citep{Santos-2001:a,Fischer-2005}. In the scenario where gas giants
acquire their mass through planetesimal coagulation followed by rapid
gas accretion onto the core, the high probability of a planet to be
present around metal-rich stars arises naturally\footnote{On the
  contrary, such a correlation is not expected with the
  gravitational-instability scenario. Recent simulations even lead to
  the opposite result \citep{Cai-2005}.} if protostellar disks attain
the same fraction of heavy material as the forming central star
\citep{Ida-2004:b}.

\begin{figure}[t]
\centering
\includegraphics[width=\hsize]{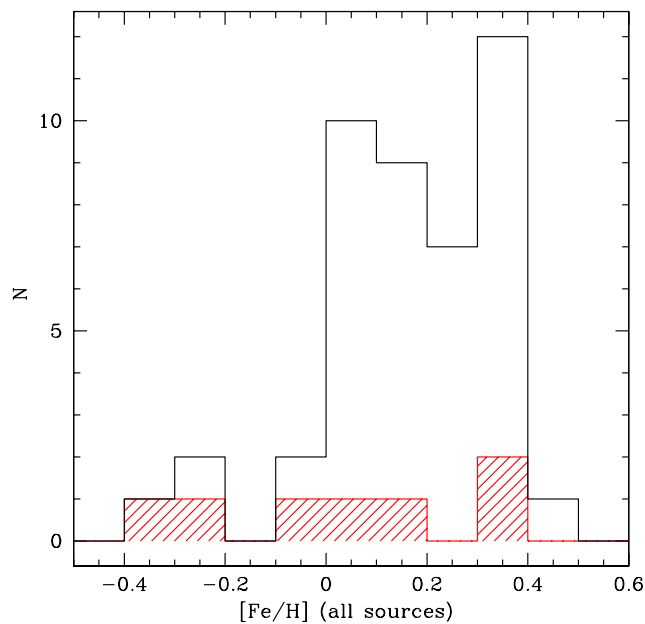}
\caption{Metallicity distribution of the sample of extrasolar planet
hosts for planets with shorter periods than 20~days. Stars with
Neptune-mass planets are indicated by the shaded histogram.}
\label{fig7}
\end{figure}

If the new {\it hot Neptune} planets are the remains of evaporated
ancient giant planets, they should also follow the metallicity trend
of their giant progenitors. This does not seem to be the case,
considering that the 7 known planets with
$m_2\sin{i}$\,$\leq$\,21\,M$_{\oplus}$ -- $\mu$\,Ara\,c
\citep{Santos-2004:a}, 55\,Cnc \citep{McArthur-2004}, Gl\,436
\citep{Butler-2004}, Gl\,777A\,c \citep{Vogt-2005}, Gl\,876\,d
\citep{Rivera-2005}, Gl\,581 \citep{Bonfils-2005}, and HD\,4308\,b --
have metallicities of 0.33, 0.35, 0.02, 0.14, $-0.03$, $-0.25$, and
$-0.31$, respectively. Although the statistics are still poor, the
spread of these values over the nearly full range of planet-host
metallicities (Fig.\,\ref{fig7}) suggests a different relation between
metal content and planet frequency for the icy/rocky planets in regard
to the giant ones.

However, we have to note here that 3 of the candidates orbit M-dwarf
primaries. Recent Monte-Carlo simulations by \citet{Ida-2005} show
that planet formation around small-mass primaries tends to form
planets with lower masses in the Uranus/Neptune domain. A similar
result that favours lower-mass planets is also observed for solar-type
stars in the case of the low metallicity of the protostellar nebula
\citep[][ Mordasini et al. in prep]{Ida-2004:b}.  Future improvements
in the planet-formation models and new detections of very-low mass
planets will help to better understand these 2 converging effects.

\section{Summary and concluding remarks}

In this paper we have reported the detection of a new very light
planet in the Uranus/Neptune mass range that orbits the deficient star
HD\,4308.  With a period of 15.56\,days ($a$\,=\,0.115\,AU), the
planet is very similar to $\mu$\,Ara\,c \citep{Santos-2004:a}.  This
new discovery was made possible by the high radial-velocity precision
reached with the HARPS spectrograph and the applied observing strategy
of averaging down the intrinsic radial-velocity ``noise'' induced by
stellar oscillation modes.  The HARPS precision is demonstrated by the
very low residuals around the orbital fit, allowing for accurate
determination of the orbital parameters.

The existence of small planets (10-100\,M$_{\oplus}$) at short and
intermediate distances from their stars strongly constrains the
standard planet formation and evolution theories. The difficulty
arises from the fast runaway accretion and the potentially large
amount of gas available for accretion, both leading towards larger
mass planets whose inward migration should turn them into hot Jupiters
(or hot Saturns). However, other effects, such as metallicity (and
HD\,4308 is metal deficient), have not yet been studied in detail.

Evaporation could possibly be invoked to account for the planets that
are very close to their star \citep{Baraffe-2004,Baraffe-2005}.
However, it is not clear whether $\mu$\,Ara\,b, and especially
HD\,4308\,b, could actually result from the evaporation of a more
massive object.  In the case of negligible evaporation or if other
small mass planets should be discovered for which evaporation can
safely be ruled out, the existence of the ``planetary desert'', or at
least its depth, must be questioned.  Therefore, from an observational
point of view, a larger sample of small objects far enough away from
the star, would be of paramount importance for further constraining
our understanding of the formation of giant planets.

Recent HARPS discoveries indicate that a population of Neptune- and
Saturn-mass planets remains to be discovered below 1\,AU. The
increasing precision of the radial-velocity surveys will help answer
this question in the near future, thereby providing us with useful new
constraints on planet formation theories.  With the precision level
achieved by HARPS, a new field in the search for extrasolar planets is
now open, allowing the detection of companions of a few Earth masses
around solar-type stars.  Very low-mass planets
($<$\,10\,M$_{\oplus}$) might be more frequent than the previously
found giant worlds.  Such planets will furthermore be preferential
targets for space missions like the photometric satellites COROT and
Kepler.

\begin{acknowledgements}
  The authors thank the different observers from the other HARPS GTO
  sub-programmes who have also measured HD\,4308. Nuno Santos is
  especially thanked for his spectroscopic analysis of the star and Yann 
  Alibert for very thoughtful comments. We would like to thank the Swiss 
  National Science Foundation (FNRS) for its continuous support of this 
  project. This study also benefitted from the support of the 
  HPRN-CT-2002-O0308 European programme.
\end{acknowledgements}

\bibliographystyle{aa}
\bibliography{udry_articles}

\end{document}